\documentclass[10pt]{article}

\usepackage{amsmath,amsthm,amssymb,color}
\usepackage{amsfonts,latexsym, hyperref, bbm, mathrsfs}
\title{\textbf{ Hawking radiation in a non-covariant frame:\\ the Jacobi metric approach}}
\author{Avijit Bera $^1$, Subir Ghosh $^1$, Bibhas Ranjan Majhi $^2$ \\
		$^1$\textit{Physics and Applied Mathematics Unit, Indian Statistical Institute}\\ 
		\textit{203 B.T. Road, Kolkata 700108, India}\\
		 $^2$\textit{Department of Physics, Indian Institute of Technology Guwahati}\\
    	\textit{Guwahati 781039, Assam, India}\\
    	{E-mails: avijitbera1997@gmail.com; subirghosh20@gmail.com; bibhas.majhi@iitg.ac.in}}
	
\begin{document}
	
	\maketitle

\begin{abstract}
The present paper deals with a reformulation of the derivation of Hawking temperature for static and stationary black holes. In contrast to the conventional approach, where the covariant form of the metrics are used, we use the manifestly non-covariant Jacobi metric for the black holes in question. In the latter, a restricted form of Hamilton-Jacobi variational principle is exploited where the energy of the particle (pertaining to Hawking radiation) appears explicitly  in the metric as a constant parameter. Our analysis shows that, as far as computation of Hawking temperature (for stationary black holes) is concerned, the Jacobi metric framework is more streamlined and yields the result with less amount of complications, (as for example, considerations of positive and negative energy modes and signature change of the metric across horizon do not play any direct role in the present analysis).
\end{abstract}
	
\section{Introduction }
	In the present article we provide yet another new approach to derive the Hawking temperature  \cite{Hawking:1974rv,Hawking:1974sw} for a black hole. Since many facts of Hawking radiation still remain unexplained it is always good to explore new avenues hoping for new insights, one such being that even a {\it manifestly non-relativistic metric} carries enough information to correctly reproduce the Hawking temperature for all known examples of static black holes. The metric in question is the Jacobi metric \cite{pin,szy,tsig,pet} that has generated a lot of interest after the pioneering work by Gibbons \cite{gib1}. Originally  Jacobi metric formalism was exploited solely in  non-relativistic Newtonian  dynamical problems \cite{pin}. The significance of Gibbon's work \cite{gib1} is that it extended the working space of Jacobi metric from non-relativistic to relativistic domain and in fact, previously known results (in the latter) were correctly recovered with less amount of computational complexity and newer insights \cite{gib2,sg}. However, it is important to emphasize that all these works primarily dealt with  particle dynamics outside of the black hole horizon. In the present paper, for the first time, we have further extended the applicability of Jacobi metric formalism in discussing  the Hawking radiation as tunneling of particles across the horizon.
	
	Hawking radiation as tunneling through horizon started from a heuristic picture, originally introduced by Hawking himself \cite{Hawking:1974sw}. This is similar to the electron-positron pair creation in a constant electric field. Here the particle and anti-particle pair is crated near to the horizon; among them one is ingoing and other one is outgoing. The ingoing mode travels inside the black hole while the outgoing candidate crosses the horizon barrier through tunneling and then ultimately observed at the infinity. Since the time and radial coordinates changes their signature inside the horizon, in this case the modes are allowed to travel along the classically forbidden path. Consequently, the allowed path becomes a complex one as inside the horizon outgoing modes is moving opposite to the forward in time.

	This heuristic picture was first formalised by Hamilton-Jacobi (HJ) approach \cite{Srinivasan:1998ty,Shankaranarayanan:2000qv} and later on by null geodesic approach \cite{Parikh:1999mf,Parikh:2004ih}. These were later developed and used to study several aspects of Hawking effect by many workers \cite{Kerner:2008qv,DiCriscienzo:2008dm,Akhmedov:2006pg,Banerjee:2008ry,Banerjee:2008cf,Banerjee:2008sn,Majhi:2009uk,Banerjee:2009wb,Banerjee:2009pf,Bhattacharya:2016kbm} (see the review \cite{Vanzo:2011wq} for a complete list of literature in this direction). However, in all these cases the relativistic form of Hamiltonian-Jacobi (HJ) variational framework was used with {\it e.g.} Schwarzschild , Reissner-Nordstrom, Kerr-Newman  metrics. The novelty  in our work is the adjustment of HJ for the (manifestly non-relativistic) Jacobi metric corresponding to each of these metrics in the context of particles tunneling out across horizon in the form  of Hawking radiation. Quite interestingly, our derivation will reveal that the all important particle energy, appearing in the thermal spectrum of Hawking radiation, is incorporated in the definition of Jacobi metric from the very beginning, whereas in the relativistic HJ one has to impose it from outside in the time space variable separation of HJ function. This makes the overall computation somewhat easier simply because time seizes to be a variable in the Jacobi metric that entirely deals with spatial degrees of freedom. Moreover, it reinforces the fact that, as far as physics is concerned, the Jacobi metric contains equivalent information like the original spacetime metric.
		
	The paper is organised as follows. In Section \ref{Jacobi} we provide a brief derivation of Jacobi metric corresponding to a generic relativistic metric. In Section \ref{SSS} we recover the Hawking temperature for static, spherically symmetric black hole in Jacobi metric approach in detail. Section \ref{Stationary} comprises calculation of Hawking temperature for stationary black holes. We conclude in Section \ref{Conclusion} with a summary of the present work and future directions.

\section{The Jacobi metric for time independent spacetimes}
\label{Jacobi}
	The principle of least action, enunciated by  Maupertuis was reframed by Jacobi where  the action of a mechanical system in phase space,
	\begin{equation}
	\label{01}
	S=\int _\gamma p_i~dx^i=\int _\gamma p_i~\dot{x}^i dt~,
	\end{equation} 
	with coordinates $x_i$ and conjugate momenta $p_i$, is varied along an unparameterized path $\gamma $ with the constraint the energy $E$ is constant along the path $\gamma $. Here dot represents the derivative with respect to time $t$. Because in general, the Jacobi metric can be curved, the solution of a Newtonian dynamical problem now reduces to geodesic motion in curved space characterized by the Jacobi metric for a given particle energy. In relativistic context, for the static space-times, it was shown in \cite{gib1}  that geodesics of a particle on an energy-dependent Riemannian metric (known as Jacobi metric) describe  the free motion of a massive particle on the corresponding spacetime. For  massless particles  this metric reduces to  the energy independent Fermat or optical metric. The structure of Jacobi metrics corresponding to  stationary (as well as non-stationary) black hole metrics are discussed in \cite{gib2}. Below we briefly describe the construction of the Jacobi metric for a general spacetime whose metric coefficients are independent of time. The details can be obtained from \cite{gib2}.
	

	Consider the following spacetime metric:
	\begin{eqnarray}
	\label{a1}
	ds^{2}=-f(x_k) dt^2+g_{ij}dx^{i}dx^{j}~.
	\end{eqnarray}	
	The action for a particle of  mass $m$ is given by	
	\begin{eqnarray}
	\label{a2}
	S=-m\int Ldt=-m\int dt\sqrt{f(x_k)-g_{ij}\dot{x^{i}}\dot{x^{j}}}~,
	\end{eqnarray}  
	where dot represents  derivative with respect to the coordinate time $t$. 
	With the canonical momenta	
	\begin{eqnarray}
	\label{a3}
	p_i=\frac{mg_{ij}\dot{x^j}}{\sqrt{f(x_k)-g_{ij}\dot{x^i}\dot{x^j}}}~,
	\end{eqnarray}
	the Hamiltonian turns out to be	
	\begin{eqnarray}
	\label{a4}
	H=\sqrt{m^2f(x_k)+f(x_k)g^{ij}p_ip_j}~.
	\end{eqnarray}
	Thus, the Hamilton-Jacobi equation becomes,	
	\begin{eqnarray}
	\label{a5}
	\sqrt{m^{2}f(x_k)+f(x_k)g^{ij}\partial_{i}S\partial_{j}S} = E~,
	\end{eqnarray}
	where  $p_i=\partial_i S$ and $E$ is the conserved energy. Rewriting (\ref{a5}) in the form	
	\begin{eqnarray}
	\label{a6}
	\frac{1}{E^{2}-m^{2}f(x_k)}f^{ij}\partial_{i}S\partial_{j}S = 1
	\end{eqnarray}
	yields  the  Hamiltonian-Jacobi equation for geodesics on the Jacobi-metric $j_{ij}$  given by	
	\begin{eqnarray}
	\label{a7}
	j_{ij}dx^{i}dx^{j} = (E^{2} - m^{2}f(x_k))f^{-1}(x_k)g_{ij}dx^{i}dx^{j}~.
	\label{B1}
	\end{eqnarray}
	Note that for massless particles ($m=0$), Jacobi metric is equivalent to the optical or Fermat metric $f_{ij}=f^{-1}(x_k)g_{ij}$, up to factor of $E^{2}$. Hence, in contrast to the massless case, for massive particles the geodesics depend upon energy E in a non-trivial way.

	
\section{Hawking temperature for a general static spherically symmetric black hole in  Jacobi Metric formalism}
\label{SSS}
	We quickly recapitulate the computational scheme for deriving Hawking temperature in the conventional Hamilton-Jacobi (HJ) approach  for the covariant (relativistic) spacetime (e.g. see \cite{Srinivasan:1998ty}). This will clearly reveal the difference between the covariant framework and spatial (manifestly no-relativistic) Jacobi metric scheme. This approach is based on semi-classical technique where the wave function for the particle is taken to be $\psi\sim e^{(i/\hbar) S}$, with $S$ is the classical HJ action. In this picture, the action is calculated by using the particle's quantum equation of motion with considering only radial motion. To find the explicit form one uses the choice of $S$ as $S(r,t) = S(r)-Et$. This choice is dictated  by the static nature of the spacetime. After finding $S$ for the specific underlying spacetime, one obtains the ingoing and outgoing probabilities, the ratio of which leads to the tunneling rate. It turns out that the rate is thermal in nature and one identifies the correct Hawking temperature by comparing with the Boltzmann factor. Below, instead of using the full spacetime metric, we will show that non-covariant Jacobi metric is sufficient to extract the correct Hawking temperature.

\subsection{Static metric in Schwarzschild coordinates}	
	Since the expressions for the Jacobi metric corresponding to the $(1+3)$-dimensional covariant metrics have already been provided in \cite{gib1,gib2,sg} we simply use these results.
	A generic static, spherically symmetric black hole in Schwarzschild coordinates take the following form:
	\begin{equation}\label{09}
	ds^2 = -f(r)dt^2 + f^{-1}(r)dr^2 + r^2(d\theta^2 + \sin^2\theta d\phi^2)~.
	\end{equation}
	The location of the horizon is determined by $f(r_H)=0$.
	The corresponding Jacobi metric, using (\ref{B1}) turns out to be \cite{gib1}, 
	\begin{equation}\label{10}
	ds^2 = j_{ij} dx^{i}dx^{j}= \Big(E^2-m^2f(r)\Big) \Big(\frac{dr^2}{f^2(r)} + \frac{r^2}{f(r)}(d\theta^2 + \sin^2\theta d\phi^2)\Big)~.
	\end{equation}
	Below we shall find the action for a particle,  moving in this background. Since the Hawking radiation is a near horizon phenomenon and tunneling occurs along the radial direction, the computation of the action will be done in this region keeping all angular coordinates fixed. This will be the main ingredient of calculating the tunneling probability of the particle through the horizon.
	
	Start with the reparametrization invariant action:
	\begin{equation}\label{11}
	S = -\int \sqrt{j_{ij}\frac{dx^i}{ds}\frac{dx^j}{ds}} ds~,
	\end{equation}
	for a particle in the background (\ref{10}).
	The integrand in the action (\ref{11}) becomes
	\begin{equation}\label{12}
	\sqrt{j_{ij}\frac{dx^i}{ds}\frac{dx^j}{ds}} = \pm(E^2-m^2f(r))^{\frac{1}{2}} (f^{-1}(r)) (\frac{dr}{ds})~.
	\end{equation}
	We shall now show that the positive and negative signs denote the  outgoing and ingoing paths, respectively. In the semi-classical picture, the wave function of the particle is given by $\psi=e^{(i/\hbar)S}$ which we shall use later. So the radial momentum of the particle comes out to be $p_r=\partial_rS$. We denote the outgoing particle which has positive momentum while the ingoing one has negative $p_r$.
Now using (\ref{12}) in (\ref{11}) one finds 
\begin{equation}\label{13}
p_r=\partial_r S= \mp(E^2-m^2f(r))^{\frac{1}{2}} (f^{-1}(r))~.
\end{equation}
 Since in our tunneling picture the particle is just inside the horizon, we must have $f(r)<0$. Then the negative sign of the above related $p_r>0$ and so it corresponds to the outgoing particle. For the same reason the other sign corresponds to ingoing one. The same argument has also been adopted in the earlier conventional tunneling approaches (e.g. see \cite{Srinivasan:1998ty}).

Now since the tunneling is a near horizon phenomena and as is well known, close to the  horizon  effectively the metric reduces to $(1+1)$-dimensions \cite{Robinson:2005pd,Iso:2006wa,biv1} with only the radial motion becoming significant.  Therefore we consider the Taylor series expansion of $f(r)$ around the horizon $r=r_H$:    	
	\begin{equation}\label{14}
	f(r) = f(r_H) + f'(r_H)(r-r_H) + \mathcal{O}(r-r_H)^2 = 2\kappa(r-r_H) + \mathcal{O}(r-r_H)^2~,
	\end{equation}
	where in second equality the expression for surface gravity of the black hole $\kappa=f'(r_H)/2$ has been used. Next substituting this in (\ref{12}) we obtain
	\begin{equation}\label{15}
	\sqrt{j_{ij}\frac{dx^i}{ds}\frac{dx^j}{ds}} = \pm \frac{E}{2\kappa(r-r_H)}(\frac{dr}{ds}) \mp \frac{m^2}{2E}(\frac{dr}{ds}) \pm \mathcal{O}(r-r_H)(\frac{dr}{ds})~.
	\end{equation}
	Hence, close  to the horizon the action for radial motion is	
		\begin{equation}\label{16}
	S = \mp \frac{E}{2\kappa}\int \frac{1}{(r-r_H)} dr \pm \frac{m^2}{2E}\int dr \mp \int \mathcal{O}(r-r_H) dr~.
	\end{equation}
For particle tunneling close to and across the horizon, the limits of  integration are taken as $r_H-\epsilon$ to $r_H+\epsilon$, with $\epsilon(>0)$  very small:
		\begin{equation}\label{17}
	S = \mp \frac{E}{2\kappa} \int_{r_H-\epsilon}^{r_H+\epsilon} \frac{1}{(r-r_H)} dr \pm \frac{m^2}{2E}\int_{r_H-\epsilon}^{r_H+\epsilon} dr \mp \int_{r_H-\epsilon}^{r_H+\epsilon} \mathcal{O}(r-r_H) dr~.
	\end{equation}
	The first integral on the right hand side of (\ref{17}), with the change of variable  $r-r_H = \epsilon e^{i\theta}$, yields \cite{Book}, 
		\begin{equation}\label{18}
	\int_{r_H-\epsilon}^{r_H+\epsilon} \frac{1}{(r-r_H)} dr = - i\pi~.
	\end{equation}
		The second integral on the r.h.s. of (\ref{17}) $\sim \epsilon \approx 0$ and	the remaining integral  is real. From hindsight we know that the real part of action is unimportant in the present context and  so,  we do not  need its exact expression. Thus the all-important form of the action is
		\begin{equation}\label{19}
	S = \pm \frac{i\pi E}{2\kappa} + \textrm{real part}~.
	\end{equation}

	Let us now distinguish the action for ingoing and outgoing trajectory  as $S_{in}$ and $S_{out}$ respectively,
		\begin{equation}\label{20}
	S_{out} = +\frac{i\pi E}{2\kappa} + \textrm{real part}, ~~~
		S_{in} = - \frac{i\pi E}{2\kappa} + \textrm{real part}~.
	\end{equation}
At semi-classical level, the WKB wave function for the particle is given by,
	$\psi = Ae^{\frac{i}{\hbar}S}$
	with $A$ being a (unimportant) normalization constant. Thus, the wave functions for outgoing and ingoing particle  $\psi_{out}$ and $\psi_{in}$ respectively are	
	\begin{equation}\label{21}
	\psi_{out} =A e^{\frac{i}{\hbar}S_{out}}, ~~~
	\psi_{in} = Ae^{\frac{i}{\hbar}S_{in}}~.
	\end{equation}
		Then the probability of the particle, coming out from the horizon is 
		\begin{equation} \label{22}
	P_{out} = |\psi_{out}|^2 = \mid A\mid ^2|e^{\frac{i}{\hbar}S_{out}}|^2 
	=\mid A\mid ^2 e^{-\frac{\pi E}{\hbar \kappa}}~.
	\end{equation}
	This also shows that the real part of the action does not contribute in the probability. In a similar way, the probability of the particle, going inside  the horizon, is
		\begin{equation} \label{23}
	P_{in} = |\psi_{in}|^2 = \mid A\mid ^2|e^{\frac{i}{\hbar}S_{in}}|^2 		
	= \mid A\mid ^2 e^{\frac{\pi E}{\hbar \kappa}}.
	\end{equation}
		Therefore the cherished expression for the tunneling rate is,
		\begin{equation}\label{24}
	\Gamma = \frac{P_{out}}{P_{in}} = e^{-\frac{2\pi E}{\hbar \kappa}} \equiv  e^{-\frac{E}{T_{H}}}
	\end{equation}
	which is identical to the Boltzmann factor with the temperature is identified as the Hawking temperature 	
	\begin{equation}\label{25}
	 T_{H} = \frac{\hbar \kappa}{2\pi}~.
	\end{equation}
 This shows that in a manifestly non-covariant framework, Jacobi metric carries sufficient amount of black hole features so that one can correctly reproduce the Hawking temperature. This demonstration constitutes one of our main results.
	
	We will conclude this section by calculating the Hawking temperature from the derived formula (\ref{25}), for some popular static black hole metrics.	
	\vskip .3cm
	{\it Example I: Schwarzschild black hole}\\
		For Schwarzschild black hole of mass $M$, the metric coefficient is given by $f(r)=(1-\frac{2M}{r})$ with horizon is at $r=2M$. Then the  surface gravity is
	\begin{equation}\label{26}
	\kappa= \frac{f'(r_H)}{2}  = \frac{1}{4M} .
	\end{equation}
	Thus we recover the well known  Hawking temperature for Schwarzschild metric,
		\begin{equation}\label{27}
	T_{BH} = \frac{\hbar}{8 \pi M}~.
	\end{equation}
	\vskip .3cm
		{\it Example II: Reissner-Nordstrom black hole}\\
		 The metric coefficient for Reissner-Nordstrom black hole with mass $M$ and charge $Q$ is
	$f(r)=(1-\frac{2M}{r} + \frac{Q^2}{r^2})$. The  
	the surface gravity is given by	
	\begin{equation}\label{28}
	\kappa= \frac{r_+-r_-}{2r_+^2}~.
	\end{equation}	
where $r_\pm = M \pm \sqrt{M^2-Q^2}$. Here $r_+$ is the event horizon we have calculated the surface gravity at this horizon. Then  by general expression (\ref{25})  one obtains the correct Hawking temperature as	
	\begin{equation}\label{29}
	T_{BH} = \frac{\hbar}{4 \pi}\frac{r_+-r_-}{r_+^2}~.
	\end{equation}
	
	So we saw that Jacobi metric also produces the correct expression for temperature through HJ quantum tunneling approach. There is a particular difference between the conventional HJ based on spacetime and the present one based on Jacobi metric. In the conventional one, as we mentioned at the beginning of this section, the quantum equation of motion (like Klein-Gordon equation for scalar particle) is used to calculate $S$ with a particular decomposition of $S$ in time and radial parts. In this case the semi-classical wave function is used at the very beginning. Whereas, here we did a completely classical calculation to obtain it by constructing an action of the particle in Jacobi metric. Later on, the quantum concept is incorporated by constructing the semi-classical wave function for the modes of the particle with the use of computed classical action. Once the wave functions are obtained, rest of steps are identical to the conventional one in finding the Hawking temperature. 
	
	The above point regarding conceptual distinction between the conventional scheme and the present Jacobi metric approach leads to another non-trivial difference. In our formalism the fermion particle tunneling is identical to boson particle tunneling. Recall that for fermion tunneling one has to consider Dirac equation (instead of Klein-Gordon equation for boson particle) and its analysis is different. However, the fermion-boson mismatch does not show up in our formalism simply because we start with the classical action that does not distinguish between fermions and bosons. In some sense the Jacobi metric approach underlines the universality of Hawking radiation.
\subsection{Static metric in Painleve coordinates}
So far we have discussed the obtention of Hawking temperature in tunnelling picture using the Jacobi metrics corresponding to static, spherically symmetric black holes, expressed in Schwarzschild coordinates. In order to show that the expression for temperature (\ref{25}) does not depend on what type of coordinates one is using to express the black hole metric, in the below, we will analyze the Jacobi metric corresponding to same black hole metrics, expressed in Painleve coordinates \cite{gib2}. Just to mention, this coordinate system is backbone of the tunneling formalism in null geodesic approach \cite{Parikh:1999mf}. Therefore here we concentrate on the black hole metrics, expressed in these coordinates.

It is well known that due to the coordinate singularity, for calculations  close or  at the horizon, the Schwarzschild form (\ref{09}) may not be always suitable. One uses a  Painleve coordinate transformation 
	\begin{equation}\label{30}
	dt \rightarrow dt-\sqrt{\frac{1-f(r)}{f^2(r)}} dr~,
	\end{equation} 
to remove the coordinate singularity. Under this the metric (\ref{09}) takes the following form:
\begin{equation}\label{31}
	ds^2 = -f(r)\Big(dt-\sqrt{\frac{1-f(r)}{f^2(r)}}dr\Big)^2 + \frac{1}{f(r)}dr^2+r^2(d\theta^2 + \sin^2\theta d\phi^2)~.
	\end{equation}

To find the Jacobi metric of the above one we start with a general discussion on finding the Jacobi metric for a more general spacetime metric of the form
\begin{equation}\label{32}
	ds^2= -v^2(x)(dt+A_idx^i)^2 + g_{ij}dx^idx^j~.
\end{equation}
Jacobi metric	for the generic form (\ref{32}) is given by \cite{gib1} 	
	\begin{equation}\label{33}
	ds^2= J_{ij}dx^idx^j = \frac{E^2 - m^2v^2(x)}{v^2(x)}g_{ij}(x)~.
	\end{equation}
	where E is the energy of the particle of mass $m$.
Now comparing (\ref{31}) and (\ref{32}) we identify
\begin{equation}\label{34}
v^2=f(r) ;~~ A_r=-\sqrt{\frac{1-f(r)}{f^2(r)}} ,~~
g_{ij}dx^idx^j = \frac{1}{f(r)}dr^2+r^2(d\theta^2 + \sin^2\theta d\phi^2)~,
\end{equation}
with all other $A_i$s vanish. With this identifications, the Jacobi metric (\ref{33}) corresponding to (\ref{31}), reduces to 
\begin{equation}\label{35}
	ds^2= J_{ij}dx^idx^j = \frac{E^2 - m^2f(r)}{f(r)}[\frac{1}{f(r)}dr^2+r^2(d\theta^2 + \sin^2\theta d\phi^2)]~.
\end{equation}
Note that the above one is identical to that (\ref{10}) obtained in Schwarzschild coordinates. Therefore proceeding similar to earlier discussion we again get the same expression (\ref{25}) for Hawking temperature.


	\section{Hawking temperature for stationary black holes in  Jacobi Metric formalism}
	\label{Stationary}
	In this section, we shall extend our discussion for stationary black holes. Using the corresponding Jacobi  metric we again will find the correct expression for temperature in tunneling formalism.

	   	 {\it Example I : Kerr black hole}\\
		For Kerr Black Hole, the metric is 
		\begin{equation}\label{36}
	ds^2 = -(1-\frac{2Mr}{\rho^2})dt^2 - \frac{4Mar\sin^2\theta}{\rho^2} dt d\phi +\frac{\rho^2}{\Delta} dr^2 + \rho^2 d\theta^2 + \frac{\sin^2\theta}{\rho^2}[(r^2+a^2)^2 -a^2\sin^2\theta]d\phi^2
	\end{equation}
		where,	$\Delta(r)= r^2-2Mr+a^2 ; ~~\rho^2(r,\theta) = r^2+a^2\cos^2\theta;~~ a=\frac{J}{M} $
	with $J$ being the angular momentum. The corresponding Jacobi metric has been found out in \cite{gib2} as 
		\begin{equation}\label{37}
	ds^2 = (E^2-m^2(1-\frac{2Mr}{\rho^2}))(1-\frac{2Mr}{\rho^2})^{-1}[\frac{\rho^2}{\Delta} dr^2 + \rho^2 d\theta^2 + \frac{\sin^2\theta}{\rho^2}[(r^2+a^2)^2 -a^2\sin^2\theta]d\phi^2]~.
	\end{equation}
		Since the tunnelling is along radial direction and if for simplicity we take $\theta =0 $ (i.e. taking particle's motion along z axis) \cite{biv1}, then the required action will be (\ref{12}) with $f(r)=(1-\frac{2Mr}{(r^2+a^2)})$. In that case $\kappa$ at the event horizon $r_H=r_+$ is given by
			\begin{equation}\label{38}
		\kappa = \frac{r_+-r_-}{2(r_+^2 + a^2)}~,
		\end{equation}
with $r_\pm = M \pm \sqrt{M^2-a^2}$. So here also we obtain the Hawking temperature of the form (\ref{25}) whose explicit form is		
		\begin{equation}\label{39}
	T_{BH} = \frac{\hbar}{4 \pi}\frac{r_+-r_-}{(r_+^2 + a^2)}.
	\end{equation}

	{\it Example II : Kerr-Newman Black Hole}\\
	The spacetime metric is 
		\begin{equation}\label{40}
	ds^2 = -f(r,\theta)dt^2+\frac{1}{g(r,\theta)} dr^2  -  2H(r,\theta)dt d\phi + \rho^2 d\theta^2 + K(r,\theta)d\phi^2
	\end{equation}
		where,	
		\begin{eqnarray}
	A_a &=& \frac{er}{\rho(r,\theta)} [(dt)_a - a^2\sin^2\theta(d\phi)_a] \nonumber\\
	f(r,\theta)&=& \frac{\Delta(r)-a^2\sin^2\theta}{\rho(r,\theta)}\nonumber\\
	g(r,\theta)&=& \frac{\Delta(r)}{\rho(r,\theta)} \nonumber\\
	H(r,\theta)&=& \frac{a\sin^2\theta(r^2+a^2-\Delta(r))}{\rho(r,\theta)}\nonumber\\
	k(r,\theta)&=& \frac{(r^2+a^2)^2-\Delta(r)a^2\sin^2\theta}{\rho(r,\theta)}\sin^2\theta \nonumber
	\end{eqnarray}
	with $\Delta(r)= r^2+a^2+e^2-2Mr ;~~ \rho^2(r,\theta) = r^2+a^2\cos^2\theta;~~ a=\frac{J}{M} $. Here
	 $J$ is the Komar angular momentum and $e$ is the charge of the blackhole. 	The corresponding Jacobi metric is \cite{gib2}  
		\begin{equation}\label{41}
	ds^2 = \frac{(E^2-m^2f(r,\theta))}{f(r,\theta)}[\frac{1}{g(r,\theta)} dr^2 + \rho^2 d\theta^2 + K(r,\theta)d\phi^2]~.
	\end{equation}
	Again for radial tunnelling	and choosing $\theta =0 $ (i.e. taking particle's motion along z axis) \cite{biv1} one identifies action as (\ref{12}) with	
	\begin{equation}\label{42}
	f(r)=g(r)=\frac{r^2-2Mr+a^2+e^2}{r^2+a^2}.
	\end{equation}
	Then the expression for Hawking temperature of the horizon $r_+$ will be again given by (\ref{25}) with $\kappa$ will be calculated from the above $f(r)$. It comes exactly identical to  the form (\ref{39}); but in this case we have $r_\pm = M \pm \sqrt{M^2-a^2-e^2}$.

		The restriction to two specific values of $\theta$ is because of the presence of the
	ergosphere. The calculation breaks down because $f(r, \theta)$ is actually negative
	elsewhere at the horizon (i.e. inside the ergosphere) and then $t$ is not properly
	timelike there . The two values $\theta = 0,\pi$ correspond to where the event
	horizon and ergosphere coincide \cite{biv1}.

	\section{Conclusion and future studies}
	\label{Conclusion}	
	To summarize, we have correctly reproduced the Hawking temperature for all known forms of static as well as stationary $3+1$-dimensional black holes  by using, in each case, the manifestly non-covariant (spatial) $3$-dimensional Jacobi metric corresponding to the covariant $3+1$-dimensional conventional forms of the metrics in question. As we have demonstrated the computations using Jacobi metric is considerably simpler. But, more significantly there are important conceptual differences as pointed out below:\\
	(i) A simplification occurs in the action functional itself. One need not consider relativistic form of Hamilton-Jacobi equation with both time and space derivatives and subsequently introduce the particle energy to make a variable separation of the action in to time and space part. The beauty of Jacobi metric approach is the the energy of the particle appears explicitly (as a parameter) in the Jacobi metric and thus the Hamilton-Jacobi equation consists of only spatial derivatives from the very beginning. The theory lives in a manifestly non-covariant space.\\
	(ii) In the covariant metric there exist complications since the interpretation of space and time-like coordinates gets interchanged as the horizon is crossed. Inside the horizon time coordinate behaves as spacelike whereas space coordinates play the role of timelike coordinates. As a result the nature of Killing vector also changes from timelike to spacelike as one crosses the horizon. This  forces one to use Kruskal time that is well behaved inside and outside the horizon. But all these complications are simply absent in the Jacobi metric formalism since the Jacobi metric is manifestly three (space) dimensional and so the question of changing over from timelike to spacelike coordinates (and vice versa) does not arise.\\
			(iii) An interesting observation is that  the Jacobi metric approach underlines the universality of Hawking radiation more strongly than the conventional covariant Hamilton-Jacobi formalism. In the latter fermions and bosons need to treated differently since they obey Dirac and Klein-Gordon equation respectively. Only after performing the analysis one finds Hawking temperature comes out the same for fermions and bosons. But this fermion-boson distinction does not appear in Jacobi metric framework since we consider a classical particle in constructing the action and only later the quantum concept is incorporated by constructing the semi-classical wave function for the modes of the particle with the use of  classical action computed earlier. 
					
			An open problem in this context is the derivation of Hawking temperature for non-stationary or time dependent metrics. Primarily this is because the Jacobi metric corresponding to a time dependent metric contains additional structure (see for example \cite{gib2}) and care is needed to treat this in a consistent way.

\end{document}